\documentclass[useAMS,usegraphicx,usenatbib,referee]{mn2e}

\title[Kinetic approaches to shock acceleration]
{Kinetic approaches to particle acceleration at cosmic ray modified shocks}

\author[E. Amato, P. Blasi and S. Gabici]
{Elena Amato$^{1}$\thanks{E-mail:amato@arcetri.astro.it},
Pasquale Blasi$^{1}$\thanks{E-mail: blasi@arcetri.astro.it},
and
Stefano Gabici$^{2}$\thanks{E-mail:Stefano.Gabici@mpi-hd.mpg.de}\\
$^{1}$INAF-Osservatorio Astrofisico di Arcetri, 
Largo E. Fermi, 5, 50125, Firenze, Italy\\
$^{2}$Max-Planck-Institut f¨ur Kernphysik, Heidelberg, Germany}

\begin{document}

\date{Accepted ----. Received -----}

%\pagerange{\pageref{firstpage}--\pageref{lastpage}} \pubyear{2007}

\maketitle

\label{firstpage}

\begin{abstract}
Kinetic approaches provide an effective description of the process of 
particle acceleration at shock fronts and allow to take into account
the dynamical reaction of the accelerated particles as well as the
amplification of the turbulent magnetic field as due to streaming
instability. The latter does in turn affect the maximum achievable
momentum and thereby the acceleration process itself, in a chain of
causality which is typical of non-linear systems. Different
kinetic approaches are characterized by different levels and types of 
approximations that also imply different computational times.
Here we present the results of two such approaches: one which is 
mathematically rigorous but rather demanding from the point of view of 
computational time, and the other which is computationally very fast 
but based on an ansatz that, while physically justified, is not rigorous.
The identification of possible differences can be crucial
in assessing the possibility of implementation of one such
calculation in hydrodynamical codes for supernova explosions. 
Special emphasis is given to a discussion of the appearance of 
multiple solutions in both approaches. 
\end{abstract}

\begin{keywords}
acceleration of particles - shock waves
\end{keywords}

\section{Introduction}

Recent observations of non-thermal radiation from shell-type supernova
remnants (SNRs) are showing that effective particle acceleration takes
place in these astrophysical objects (see \cite{funk} for a recent
review). In particular the detection of spatially resolved non-thermal
X-rays from thin regions close to the forward shock in several SNRs
has showed the first clear evidence for strong magnetic field
amplification as could be expected if the shock is efficiently
accelerating cosmic rays. The recent detection of TeV gamma rays
(\cite{HESSRXJ1713,HESSRXJ1713_II,HESSRXJ1713_III,HESSRXJ0852,HESSRXJ0852_II})
is likely to provide further information on the magnetic field
amplification by adding to multifrequency observations and
by possibly allowing the measurement of the shape of the TeV gamma ray
spectrum.  

In the scenario with large magnetic field, low fluxes of gamma
radiation through inverse Compton scattering (ICS) of electrons are
expected, thereby leaving a hadronic origin of the gamma ray emission 
as a more likely possibility. This latter conclusion is however of more
limited strength and requires further confirmation through a
continuous effort to detect SNRs in gamma rays, not only in the TeV
region but also in the GeV energy range, as should become possible
after the upcoming launch of the GLAST gamma ray telescope.  

X-rays observed in the 1-10 keV energy range from SNRs are generated
by synchrotron emission of relativistic electrons and the thickness of
the brightness profiles allows one to estimate the total field in the
shock vicinity to be of order 100-500 $\mu G$ in basically all cases
in which measurements exist (\cite{volkSNR}). 
This appears to be the strongest evidence
for efficient cosmic ray acceleration at SNR shocks, where here we use
the word {\it efficient} to indicate that an appreciable fraction of
the kinetic pressure at the shock is transformed into accelerated
particles, which in turn change the dynamics of the shock. In this
regime the standard {\it test particle theory} (TPT) fails and a
non-linear theory is required (see \cite{maldru} for a review). 
From the phenomenological point of view the introduction of the
non-linear effects is crucial.

The reaction of the accelerated particles onto the shock structure has
been calculated within different approaches, both semi--analytical
(see \cite{maldru} for a recent review) and numerical
(e.g. \cite{don,kang} and references therein). 
Semi-analytical kinetic approaches provide a detailed description of
the process of particle acceleration at cosmic ray modified shocks,
including particle spectra, modified shock dynamics and thermodynamics,
and have recently been generalized to include magnetic field amplification
(\cite{amato1,amato2}). 

The main limitation of these approaches is the fact that they all
solve the stationary problem of acceleration. This problem is simply
ill defined in the case of acceleration of protons (or nuclei) in
sources such as supernova remnants: in fact in these cases there is no
appreciable energy loss process, and stationarity cannot be
reached. This is in principle true even in test particle approaches,
but there the problem is limited to momenta close to the maximum
momentum, while the shape of the spectrum at lower momenta is rather
well defined and stationary. In the non-linear regime, the entire particle
spectrum changes as a consequence of the time dependence of the
maximum momentum, therefore the assumption of stationarity is in
principle not well justified, though it is tacitly assumed that the
situation to describe is one of quasi-stationarity. 

Several approaches to time-dependent shock acceleration at cosmic ray
modified shocks have been put forward (\cite{bell87,FalGid,kang}), but
all of them are based on numerical approaches (e.g. finite difference
solution of the equations). In particular, \cite{FalGid} found that
quasi-stationary solutions are rather good approximations to the full
time-dependent solutions. Despite this, a generalization of the
semi-analytical approaches to include time dependence would be highly
desirable, and would probably help to shed some light on the problem
of the appearance of multiple solutions, as discussed below.

The first kinetic model that provided a semi-analytical (stationary)
solution of the problem was developed by \cite{malkov1} and
\cite{malkov2} for strongly modified shocks with a given, spatially constant, 
diffusion coefficient. A simple, approximate kinetic approach
was later proposed by \cite{blasi1,blasi2}. This approach provides an
accurate description of the shock modification and its effects on
particle acceleration provided the diffusion coefficient has a
sufficiently strong dependence on momentum $p$ (the underlying
assumption is somewhat similar to the starting assumption adopted by
\cite{eichler79}). 

It also allows to obtain the solution to the problem in a very short
computational time, which is particularly important when the particle
acceleration process must be described in the context of complex and
time consuming hydrodynamical simulations (for instance in the case of
SNRs). \cite{vannoni} showed, among other things, that the solutions
obtained with Malkov's method and those of Blasi's method are in very
good agreement for Bohm diffusion coefficient, while the agreement
becomes gradually worse for Kraichnan ($D(E)\propto E^{1/2}$) and 
Kolmogorov ($D(E)\propto E^{1/3}$) diffusion coefficients, as
expected. Even in these two cases, however, the differences were not 
very large and the model of \cite{blasi1,blasi2}
still provided an acceptable description of the main physical aspects
of the problem.  

More recently a solution of the system of equations describing
the transport of accelerated particles and the dynamics of the shock
was found by \cite{amato1}, where the diffusion coefficient could 
formally be taken as an arbitrary function of both momentum and spatial
coordinate (although solutions were actually computed for a few specific 
choices of $D(x,p)$). Contrary to the simple approach of \cite{blasi1} where a
physically reasonable ansatz was used to simplify the problem, this 
approach is mathematically rigorous but more expensive from the 
computational point of view.

In a second paper by \cite{amato2} the self-generation of the waves and
the determination of the resulting diffusion coefficient, though in
the context of quasi-linear theory, were introduced. The method for 
taking into account magnetic field amplification during the acceleration 
process at a modified shock, as assessed in this work, also
allowed for the determination of the maximum momentum in these complex
circumstances (\cite{caprioli}).  

Since the two methods introduced above aim at describing the same
problem but using a different set of approximations and, very important,
with a quite different computational cost, we wish to assess here
the advantages and disadvantages of using one or the other. While
doing so, we also discuss an important feature which is found
in both approaches, and more generally in semi-analytical kinetic
approaches (as well as in stationary two-fluid models
\cite{dr_v80,dr_v81}), namely the appearance of multiple solutions.

We conclude that for most applications related to the description of
the phenomenology of particle acceleration in supernova remnants the
model of \cite{blasi1,blasi2} can be succesfully applied despite its 
simplifications, provided the diffusion coefficient scales rapidly 
enough with momentum. This is certainly the case for Bohm diffusion.  
Moreover we complete this simple model with recipes that allow us to 
determine the level of magnetic field amplification and the maximum 
energy of the accelerated particles.  

The present paper is structured as follows: in \S
\ref{sec:twoapproaches} we briefly introduce the two kinetic
approaches of \cite{blasi1}, model A hereafter, and \cite{amato1}, model B. 
In the subsections we discuss in detail the results of the
two theoretical approaches. In \S \ref{sec:pmax} we describe the
introduction in the first approach of a recipe for the acceleration 
time which allows us to
estimate the maximum momentum of the accelerated particles. We
conclude in \S \ref{sec:concl}.   

\section{Two kinetic approaches}
\label{sec:twoapproaches}

In this section we briefly discuss two kinetic approaches, the
one of \cite{blasi1,blasi2} (Model A) that is expected to work for
diffusion coefficients $D(p)$ that change rapidly enough with the
particle momentum, and the one proposed by \cite{amato1} (Model B) and 
valid for any choice of $D(p,x)$. The main point of the section is to
illustrate the results obtained with the two approaches, identifying
the advantages and possible shortcomings of the first, which is being
implemented in some computationally heavy hydrodynamical codes, in
which the formal solution of Model B would be hardly possible to
introduce due to the much longer computational time required. 

The basic equations are the transport equation:
\begin{equation}
\frac{\partial}{\partial x}
\left[ D(x,p)  \frac{\partial}{\partial x} f(x,p) \right] - 
u  \frac{\partial f (x,p)}{\partial x} + 
\frac{1}{3} \left(\frac{d u}{d x}\right)
~p~\frac{\partial f(x,p)}{\partial p} + Q(x,p) = 0,
\label{eq:trans}
\end{equation}
and the conservation equation for the total momentum:
\begin{equation}
\xi_c (x) = 1 + \frac{1}{\gamma_g M_0^2} - U(x) - \frac{1}{\gamma_g M_0^2}
U(x)^{-\gamma_g}.
\label{eq:normalized1}
\end{equation}
Here $f(x,p)$ is the particle distribution function in the shock
frame, $u(x)$ is the velocity of the background fluid, which equals
$u_2$ downstream and changes continuously upstream, from $u_1$
immediately upstream of the subshock to $u_0$ at upstream
infinity. The quantity $U(x)=u(x)/u_0$ is the normalized velocity,
bound to equal unity at $x\to -\infty$. In Eq. \ref{eq:normalized1},
$\xi_c(x)=P_{CR}(x)/\rho_0 u_0^2$ is the cosmic ray pressure at the
position $x$, normalized to the kinetic pressure $\rho_0 u_0^2$ at
upstream infinity. It is customary to introduce the compression factor
$R_{sub}=u_1/u_2$ at the subshock and the total compression factor
$R_{tot}=u_0/u_2$. Assuming homogenisation of the cosmic ray
plasma in the downstream section ($df/dx^-=0$, a consequence of the
assumption of stationarity) one easily obtains (\cite{blasi1})
that the distribution function of the particles at the shock location 
is  
\begin{equation}
f_0 (p) = \left(\frac{3 R_{tot}}{R_{tot} U_p(p) - 1}\right) 
\frac{\eta n_0}{4\pi p_{inj}^3} 
\exp \left\{-\int_{p_{inj}}^p 
\frac{dp'}{p'} \frac{3R_{tot}U_p(p')}{R_{tot} U_p(p') - 1}\right\},
\label{eq:inje}
\end{equation}
where we introduced the function $U_p(p)=u_p/u_0$, with
\begin{equation}
u_p = u_1 - \frac{1}{f_0(p)} 
\int_{-\infty}^0 dx (du/dx)f(x,p)\ .
\label{eq:up}
\end{equation}
This result is very general and is not related to the assumptions
  of the model of \cite{blasi1}.
Here we assumed, as usual, that the injection is a delta function in
space (at the shock) and in momentum (at the injection momentum
$p_{inj}$) in the form $Q(x,p) = \frac{\eta n_{gas,1} u_1}{4\pi
  p_{inj}^2} \delta(p-p_{inj})\delta(x)$, with $n_{gas,1}=n_0
R_{tot}/R_{sub}$ the gas density immediately upstream ($x=0^-$) and
$\eta$ the fraction of the particles crossing the shock which are
going to take part in the acceleration process.

The compression factors $R_{sub}$ and $R_{tot}$, if the gas in the
upstream frame evolves adiabatically, are related through the
following expression:
\begin{equation}
R_{tot} = M_0^{\frac{2}{\gamma_g+1}} \left[ 
\frac{(\gamma_g+1)R_{sub}^{\gamma_g} - (\gamma_g-1)R_{sub}^{\gamma_g+1}}{2}
\right]^{\frac{1}{\gamma_g+1}}.
\label{eq:Rsub_Rtot}
\end{equation}

\vskip 1cm
{\it Model A}
\vskip .5cm
This model was proposed by \cite{blasi1} and put in a form to include
the reacceleration of seed particles by \cite{blasi2}. It is
characterized by a simple implementation that makes it fast in terms of
numerical computation while keeping all the main physical ingredients
of the problem. For these reasons it has been implemented in
hydrodynamical codes in order to calculate the instantaneous spectrum
of accelerated particles in an expanding supernova shell
(\cite{noiSNR}). 

The essence of the approach is based on the observation that the
function $f(x,p)$ in Eq. \ref{eq:up} is expected to suffer an
exponential suppression at a distance $x_p$ such that $\int_{x_p}^0 dx
u(x)/D(p)\sim 1$. The spatial position where this suppression is found
in general depends on momentum if the diffusion coefficient depends on
momentum. These spatial locations for different values of $p$ are well
defined if $D(p)$ is a strong function of momentum, as was first
pointed out by \cite{eichler79}, and become more ill defined for
$D(p)$ weakly dependent on $p$. As a first approximation
we may assume that the distribution function is $f(x,p)=f_0(p)$ at $|x|\leq
|x_p|$ and vanishes at $|x|>|x_p|$. With this assumption Eq. \ref{eq:up}
gives $u_p\approx u(x_p)$. The distance $x_p$ has the meaning of the
typical distance to which particles with momentum $p$ can diffuse away
from the shock in the upstream fluid, and the fluid velocity at that
point is $u(x_p)\approx u_p$. It follows that Eq. \ref{eq:normalized1}
can be transformed into an equation which depends only on the particle
momentum $p$:
\begin{equation}
\xi_c (p) = 1 + \frac{1}{\gamma_g M_0^2} - U_p - \frac{1}{\gamma_g M_0^2}
U_p^{-\gamma_g},
\label{eq:normalized2}
\end{equation}
where 
\begin{equation}
\xi_c (p) \approx \frac{4\pi}{3\rho_0 u_0^2} 
\int_p^{p_{max}} dp p^3 v(p) f_0(p),
\end{equation}
and $v(p)$ is the velocity of particles with momentum
$p$. Differentiating Eq. \ref{eq:normalized2} with respect to momentum
we finally have:
\begin{equation}
p\frac{d U_p}{d p} \left[ 1 - \frac{1}{M_0^2} U_p^{-(\gamma_g+1)}
  \right] = \frac{4\pi}{3 \rho_0 u_0^2} p^4 v(p) f_0(p).
\label{eq:differential}
\end{equation}
For $p\to p_{inj}$ we have $U_p\to u_1/u_0=R_{sub}/R_{tot}$, while for
$p\to p_{max}$ one has $U_p\to 1$. The procedure to calculate the
solution is therefore straightforward: for a given guess of $R_{sub}$
(and therefore of $R_{tot}$ through Eq. \ref{eq:Rsub_Rtot}) we can
solve the differential equation Eq. \ref{eq:differential} coupled with
the expression for $f_0(p)$ as a function of $U_p$ (Eq. \ref{eq:inje})
in an iterative
way. In general we end up with $U_p(p_{max})\neq 1$, which implies that 
the given value of $R_{sub}$ is not a solution of our problem. The
solution corresponds to that value of $R_{sub}$ (and $R_{tot}$) for
which $U_p(p_{max})=1$. This value also identifies completely the
function $U_p$ and the distribution function at the shock $f_0(p)$,
which is the sought after result. It is important to notice that this
procedure does not lead only to the determination of $f_0(p)$ but also
to the values of the thermodynamical quantities at the shock, such as
the temperature of the gas upstream and downstream. 
What Model A cannot provide is the spatial dependence of the
quantities in the precursor, although we provide below a physical
ansatz that may indirectly make this piece of information available. We later
check the results by comparing them with the outcome of Model B.

In both approaches that we describe here we adopt the injection
recipe known as {\it thermal leakage} (\cite{vannoni,gieseler})
which connects the value of $\eta$ to that of $R_{sub}$ in a unique
way through the relation: 
\begin{equation}
\eta = \frac{4}{3\pi^{1/2}} (R_{sub}-1) \xi^3 e^{-\xi^2}.
\end{equation}
Here $\xi$ is a parameter that identifies the injection 
momentum as a multiple of the momentum of the thermal particles in
the downstream section ($p_{inj}=\xi p_{th,2}$). The latter is an
output of the non-linear calculation, since we solve exactly the modified
Rankine-Hugoniot relations together with the cosmic rays' transport 
equation. For the numerical calculations that follow we typically use 
values of $\xi$ between $3$ and $4$, which correspond to fractions 
of order $\sim 10^{-4}-10^{-5}$ of the particles crossing the shock 
to be injected in the accelerator. 

\vskip 1cm
{\it Model B}
\vskip .5cm

A formal solution of the transport equation and the conservation
equations for an arbitrary choice of the diffusion coefficient in both
its dependence on momentum and spatial coordinate was found by
\cite{amato1}. 

\cite{malkov1} and \cite{amato1} showed that an excellent
approximation to the solution $f(x,p)$ has the form 
\begin{equation}
f(x,p) = f_0(p) \exp\left[-\frac{q(p)}{3}(1-\frac{u_1}{u_2})
\int_x^0 dx' \frac{u(x')}{D(x',p)}
\right],
\label{eq:solution}
\end{equation}
where $q(p)=-\frac{d\ln f_0(p)}{d \ln p}$ is the local slope of
$f_0(p)$ in momentum space. In fact the factor $1-\frac{u_1}{u_2}$ in
the argument of the exponential was introduced by \cite{caprioli} in
order to satisfy exactly the boundary condition at the shock, even for
weakly modified shocks. 

In terms of the distribution function (Eq.~\ref{eq:solution}), we can 
also write the normalized pressure in accelerated particles as:
\begin{equation}
\xi_c (x) = \frac{4\pi}{3\rho_0 u_0^2} \int_{p_{inj}}^{p_{max}} dp\ p^3
v(p) f_0(p) \exp\left[ -\int_x^0 dx' \frac{U(x')}{x_p(x',p)}
\right],
\label{eq:normalized3}
\end{equation}
where for simplicity we introduced $x_p(x,p)=\frac{3D(p,x)}{q(p) u_0}$.

By differentiating Eq.~\ref{eq:normalized3} with respect to $x$ we obtain
\begin{equation}
\frac{d\xi_c}{dx} = \lambda(x) \xi_c(x) U(x),
\label{eq:differ}
\end{equation}
where
\begin{equation}
\lambda(x)=<1/x_p>_{\xi_c}=
\frac{\int_{p_{inj}}^{p_{max}} dp~p^3 \frac{1}{x_p(x,p)} v(p) f_0(p) 
\exp\left[ -\int_x^0 dx' \frac{U(x')}{x_p(x',p)}\right]}
{\int_{p_{inj}}^{p_{max}} 
dp~p^3 v(p) f_0(p) \exp\left[ -\int_x^0 dx'
  \frac{U(x')}{x_p(x',p)}\right]}, 
\label{eq:lambda}
\end{equation}
and $U(x)$ is expressed as a function of $\xi_c(x)$ through 
Eq.~\ref{eq:normalized1}. 

Finally, after integration by parts of Eq.~\ref{eq:up}, one is
able to express $U_p(p)$ in terms of an integral involving $U(x)$
alone: 
\begin{equation}
U_p(p) = \int_{-\infty}^0 dx\ U(x)^2 \frac{1}{x_p(x,p)}
\exp\left[ -\int_x^0 dx' \frac{U(x')}{x_p(x',p)}\right]\ ,
\label{eq:up2}
\end{equation}
which allows one to easily calculate $f_0(p)$ through Eq.~\ref{eq:inje}. 

Eqs.~\ref{eq:normalized1} and \ref{eq:differ} can be solved by iteration 
in the following way: for a fixed value of the 
compression factor at the subshock, $R_{sub}$, the value of the
dimensionless velocity at the shock is calculated as $U(0)=R_{sub}/
R_{tot}$. The corresponding pressure in the form of accelerated
particles is given by Eq.~\ref{eq:normalized1} as 
$\xi_{c}(0) = 1 + \frac{1}{\gamma_g M_0^2} -\frac{R_{sub}}{R_{tot}}
- \frac{1}{\gamma_g M_0^2} \left(\frac{R_{sub}}{R_{tot}}\right)^{-\gamma_g}$.
This is used as a boundary condition for Eq.~\ref{eq:differ}, where
the functions $U(x)$ and $\lambda(x)$ (and therefore $f_0(p)$) on the 
right hand side at the $k^{th}$ step of iteration are taken as the
functions at the step $(k-1)$. In this way the solution of 
Eq.~\ref{eq:normalized1} at the step $k$ is simply
\begin{equation}
\xi_c^{(k)}(x) = \xi_c(0) \exp\left[-\int_x^0 d x' 
\lambda^{(k-1)}(x') U^{(k-1)}(x')\right],
\end{equation}
with the correct limits when $x\to 0$ and $x\to -\infty$. At each 
step of iteration the functions $U(x)$, $f_0(p)$, $\lambda(x)$ are
recalculated (through Eq.~\ref{eq:normalized1}, 
Eqs.~\ref{eq:up2} and \ref{eq:inje}, and Eq.~\ref{eq:lambda}, respectively), 
until convergence is reached. The solution of this set of equations, 
however, is also a solution of our physical problem only if the pressure
in the form of accelerated particles as given by Eq.~\ref{eq:normalized1} 
coincides with that calculated by using the final $f_0(p)$ in 
Eq.~\ref{eq:normalized2}. This occurs for a value of $R_{sub}$, which
fully determines the solution of our problem for an arbitrary
diffusion coefficient as a function of location and momentum. 

\subsection{Spectra and velocity profiles}
\label{sec:spectra}

\begin{figure}
\resizebox{\hsize}{!}{
\includegraphics{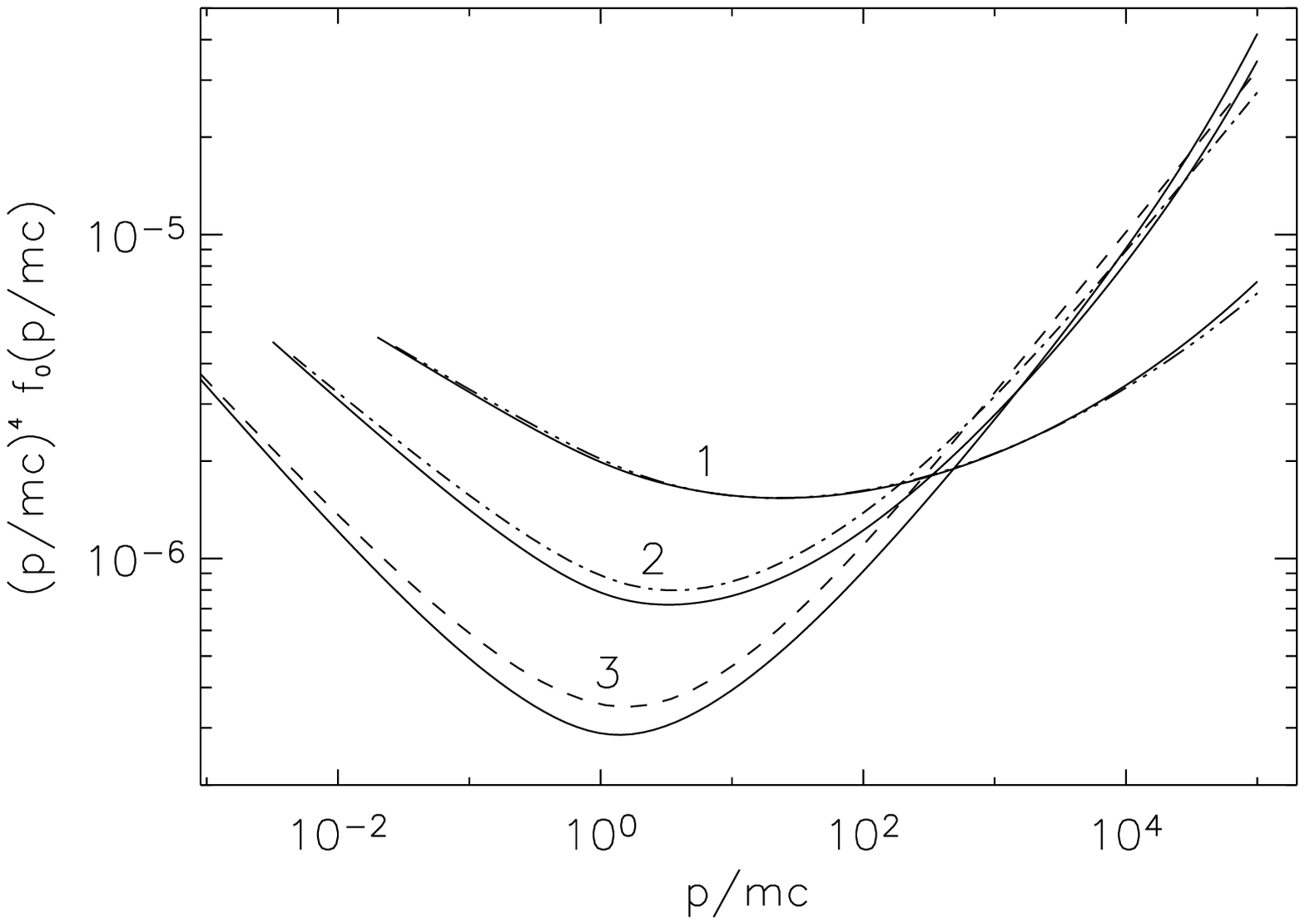}
\includegraphics{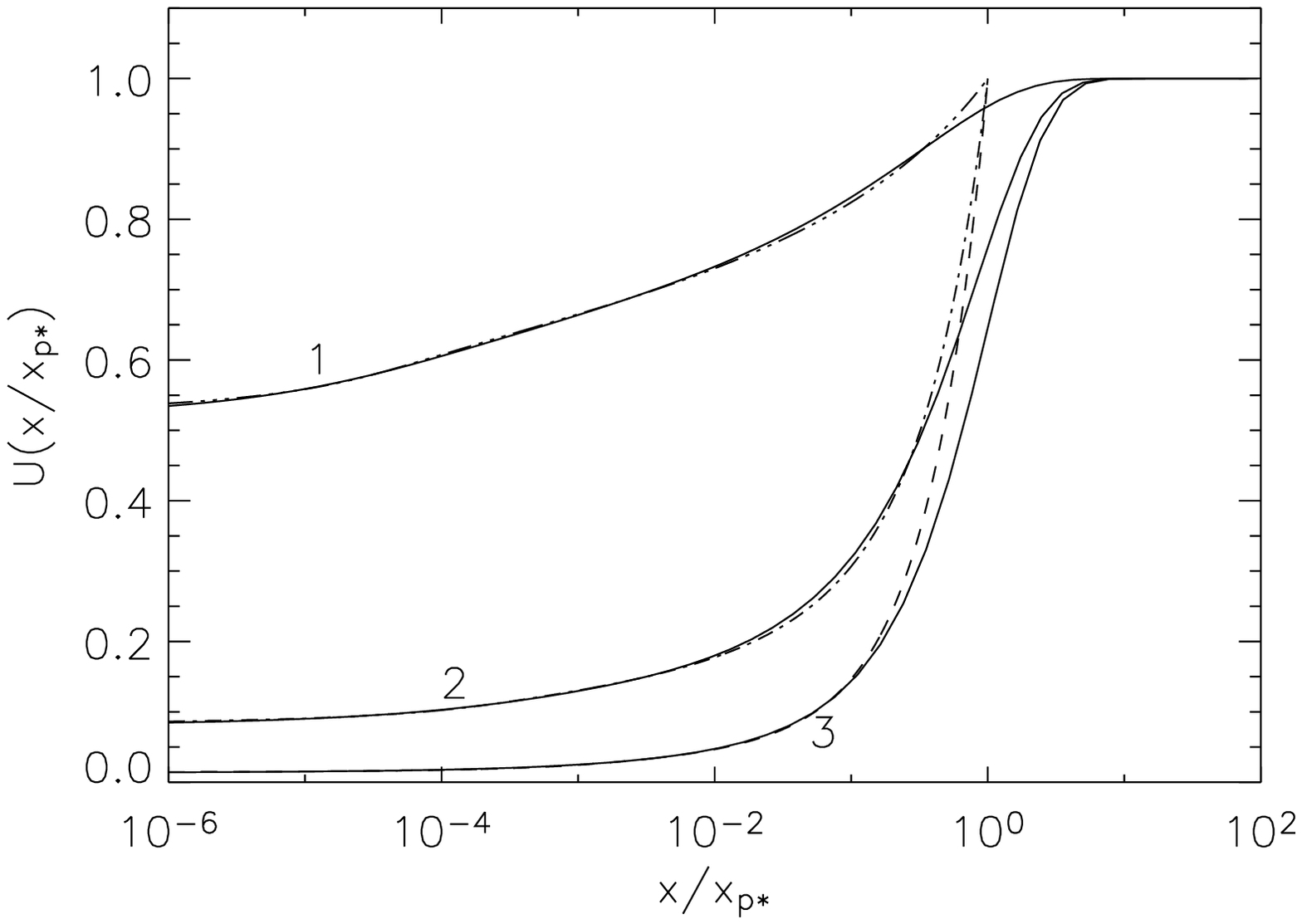}}
\caption{{\it Left Panel:} 
  Spectra of accelerated particles for
  $p_{max}=10^5 m_p c$, $u_0=5\times 10^8\rm cm~s^{-1}$, $\xi=3.5$  
  and for $M_0=10,\,100,\,1000$ (curves labeled as 1, 2 and 3 respectively).
  Bohm diffusion coefficient is adopted. The solid lines are obtained with the calculation of \citet{amato1} (model B), the dashed ones with that of \citet{blasi1} (model A). 
  {\it Right Panel:} 
  Velocity profiles in the precursor for the cases in the left panel. Again the
solid curves are computed with Model B and the dashed ones with Model A. The spatial coordinate is in units of $x_{p*}$, with $|x_{p*}|=D(p_{max})/(u_0 U_p(p_{max}))$ as discussed in the text.}  

\label{fig:spectrum}
\end{figure}

All approaches to particle acceleration at modified shocks predict the
formation of a precursor in the upstream region, resulting in a
gradient of the velocity profile of the fluid. Since qualitatively the
spectrum of the accelerated particles is still determined by an
effective compression factor felt by the particles of given momentum,
and the velocity in the precursor increases with the distance from the
shock, it is easy to infer that the spectrum of the accelerated
particles is not expected to be a
power law and more precisely that it should be concave (steeper at low
energies and flatter at high energies). Here we discuss the detailed 
shape of the spectrum at the shock as obtained through the two kinetic 
approaches described above. In Fig. \ref{fig:spectrum} (left panel) we
plot the spectra as a function of the momentum of particles for 
model B (solid lines) and for model A (dashed lines). The curves are
obtained for 
$p_{max}=10^5 m_p c$, $u_0=5\times 10^8\rm cm~s^{-1}$, $\xi=3.5$ 
and for the values of the Mach number at upstream infinity 
$M_0=10,\,100,\,1000$ (curves labeled as 1, 2 and 3, respectively). 
Bohm diffusion is assumed. The agreement between the two sets of curves 
is excellent for relatively low Mach numbers ($M_0\sim 10$) and remains 
good even up to
much larger Mach numbers, and in fact for all values we have tried. The 
largest discrepancies between the two methods are at the level of
$\sim 20\%$. The reason for such discrepancies is to be found in the
assumption that $U_p(p_{max})$ is required to equal unity in the
approach of Model A. 

The velocity profile of the fluid in the precursor is plotted in
Fig.~\ref{fig:spectrum} (right panel) for the two models (again solid
and dashed lines respectively).  

On the x-axis we plot the distance $x$ from the shock in the upstream 
region in units of $x_{p*}$, which is defined as 
$$
|x_{p*}| = \frac{D(p_{max})}{u_0 U_p(p_{max})}.
$$
Some comments are required on the calculation of $U(x)$ for Model A.
As discussed in the previous section, this
model does not keep any information about the spatial dependence of
the quantities in the precursor, although such information is somehow
contained in the relation between a momentum $p$ and the mean
diffusion length of particles with such momentum, $|x(p)| \approx
D(p)/u_p(p)$. The dashed lines in Fig.~\ref{fig:spectrum} (right
panel) are obtained in the following way: for a given location $x$
upstream, the equation $x = D(p)/u_p(p)$ is inverted and a
corresponding value $p$ of the minimum momentum of particles that
may have diffused to the point $x$ is obtained. At this point the
velocity $U(x)$ (in units of $u_0$) is by definition $U_p(p)$ for the
value of $p$ corresponding to $x$. 
By definition the fluid velocity in the simple model is bound to be 
unity at $x/x_p=1$ because no particles are supposed to be
able to reach farther regions. In the exact solution there is a spread
in the distances that can be diffusively reached at given momentum and
the transition to $U(x)=1$ is smoother. This difference in the
velocity profile affects mainly the results for the spectrum at $p\sim
p_{max}$, but since these particles carry an appreciable amount of
energy in the case of modified shocks, the whole spectral shape is
somewhat affected (at the level of at most $\sim 20\%$ in the strongly
modified cases). 

As discussed above, the ingredient of model A that makes it
much faster in terms of computational time consists in the assumption
that particles with given momentum all travel to some given maximum
distance from the shock in the upstream region. This physical ansatz
was first put forward by \cite{eichler79}. This assumption has
the precious implication that the calculation of the spectrum of
accelerated particles, of the velocity profile in the precursor, and of
all thermodynamical properties of the background plasma, can be carried
out without knowing {\it a priori} the diffusion coefficient as a
function of momentum, which in general is an input to the problem and
is very poorly known. The solid curves plotted in Figs.~\ref{fig:spectrum}
are obtained for a Bohm diffusion coefficient in Model B,  
and as stressed above the agreement between the two 
approaches is very good. As discussed by \cite{vannoni} the goodness
of the result becomes gradually worse for diffusion coefficients in
the form of Kraichnan ($D(p)\propto p^{1/2}$) and Kolmogorov
($D(p)\propto p^{1/3}$) that show a weaker dependence on momentum. 
On the other hand, X-ray observations of SNRs shocks seem to hint to
Bohm-like diffusion coefficients (see \cite{stage2006}
for recent results). 

The price to pay for the short computational time is that one can keep
track of the spatial dependence of the different quantities involved
only through the approximate relation $x\approx - D(p)/u_p(p)$. In
Model B, which is more challenging computationally, the spatial
dependence is fully accounted for and any form of the diffusion
coefficient can be adopted. This appears to be especially important
since the diffusion coefficient as calculated by \cite{amato2} in the
case of classical streaming instability leading to strong magnetic
field amplification has a flat dependence on momentum in the energy
region close to the maximum momentum (although the overall shape and
normalization of the diffusion coefficient are close to Bohm-like).    

\subsection{Compression factors and multiple solutions}
\label{sec:multiple}

The stratification of the cosmic ray pressure in the upstream region,
together with the escape of particles with momentum $p_{max}$ from
upstream infinity (equivalent to having a radiative shock), make the
fluid more compressible and therefore lead to an increase of the total
compression factor $R_{tot}$ between upstream infinity and
downstream. In this 
section we investigate the behavior of $R_{tot}$ as a function of some 
parameters of the problem, most notably the parameter $\xi$ which
defines the injection momentum and the Mach number $M_0$. 

\begin{figure}
\resizebox{\hsize}{!}{
\includegraphics{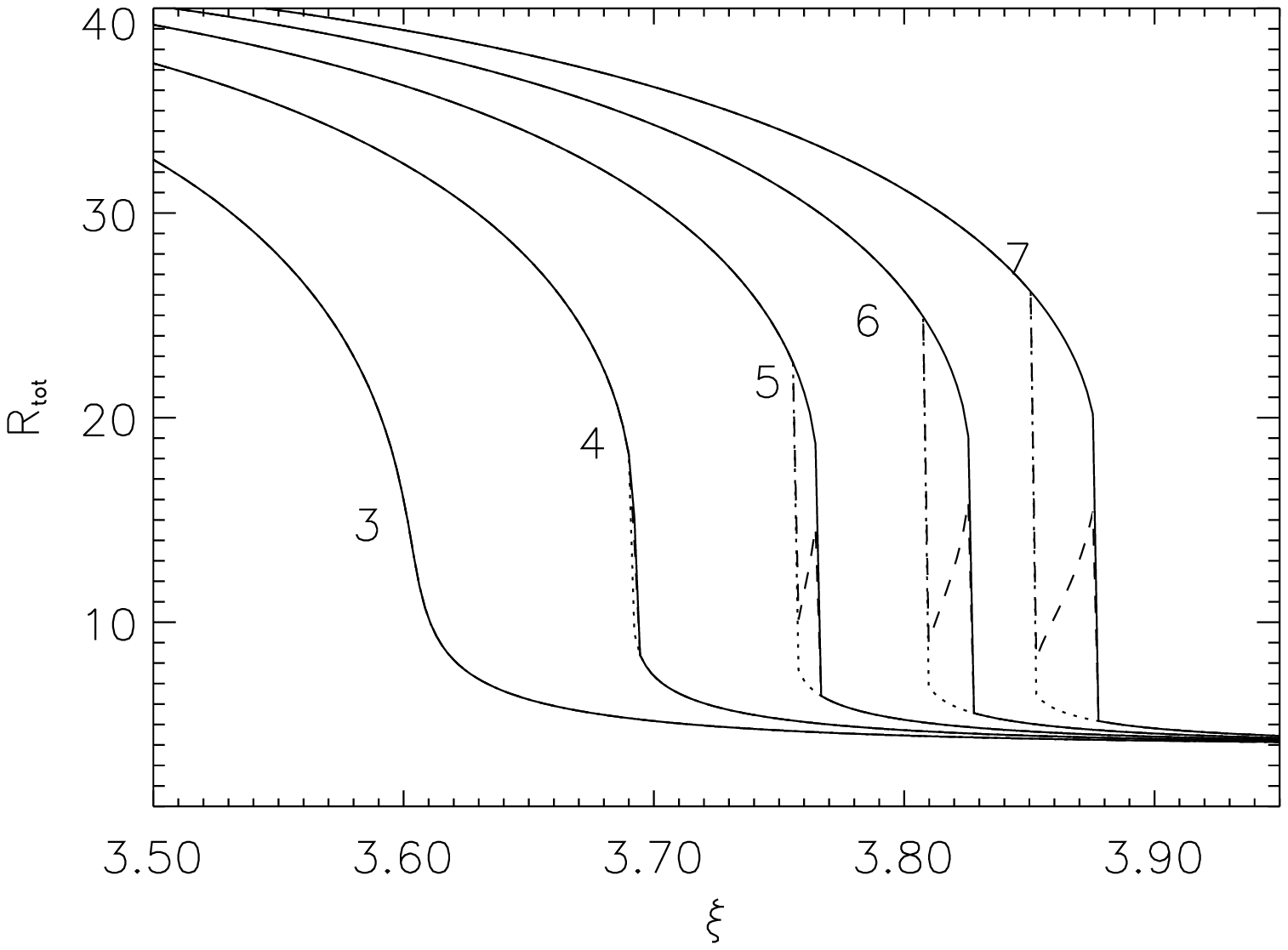}
\includegraphics{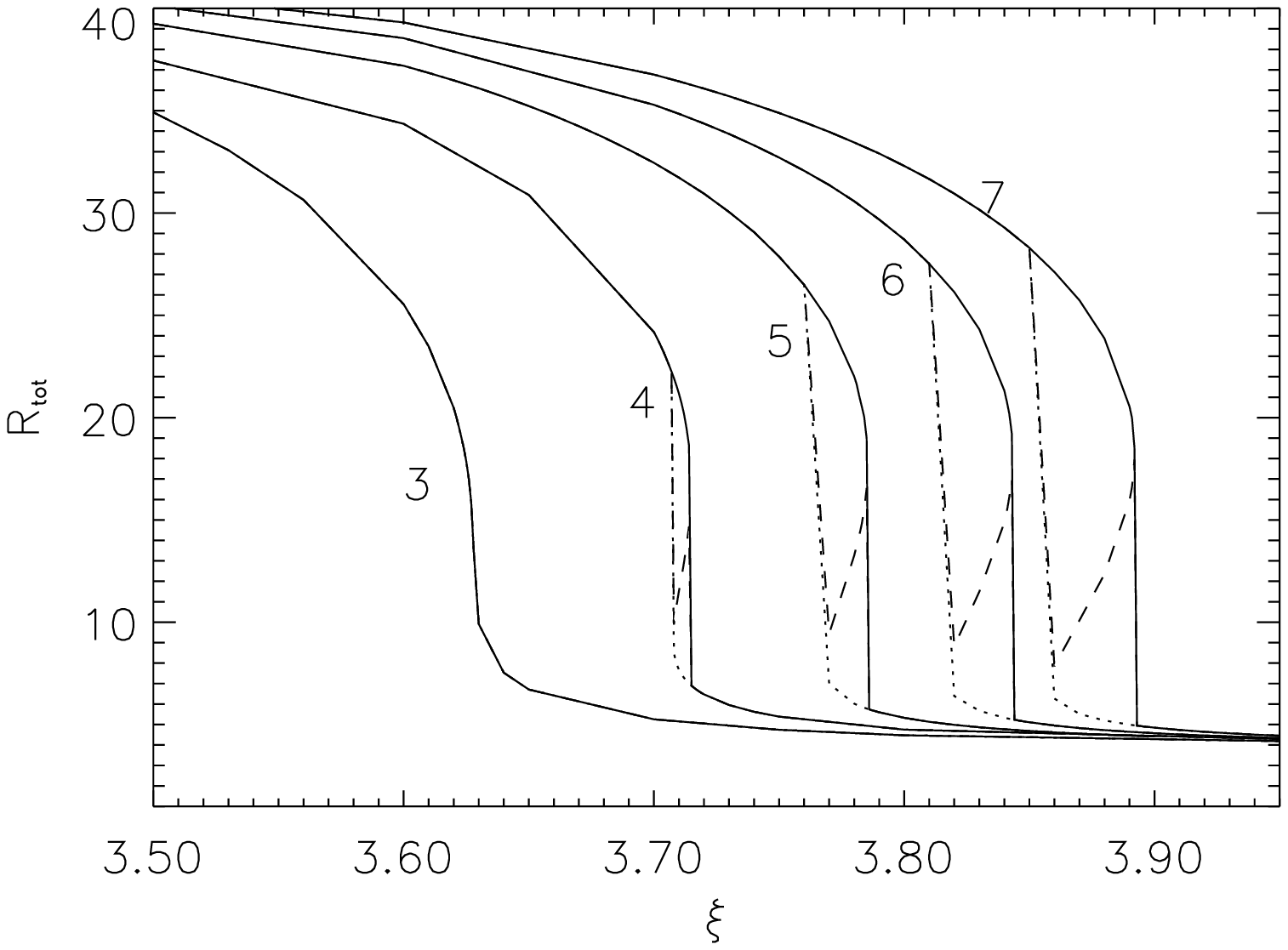}}
\caption{{\it Left Panel:} Total compression factor as a function of
  the parameter $\xi$ as obtained with the approach of
  \citet{blasi1}. The curves are labelled by the value of 
$\log_{10}(p_{max}/m_p c)$ adopted. Where multiple solutions are present, the 
solid line traces the most modified one, the dotted line refers to the one 
close to linear, and
the dashed line to the intermediate one (see text for discussion).
{\it Right Panel:} The same curves as obtained with the
  approach of \citet{amato1}. Both plots are obtained for $M_0=100$,
  $u_0=5\times 10^8\rm cm~s^{-1}$. }
\label{fig:rtotxi}
\end{figure}

The fraction of particles that are injected and take part in the
acceleration process increases when $\xi$ decreases and as a
consequence the compression factor also increases. In
Fig. \ref{fig:rtotxi} (left panel) we plot the total compression
factor as obtained in Model A. The same quantity for the
Model B is plotted in the right panel of
Fig. \ref{fig:rtotxi}. Both plots are obtained for $M_0=100$,
$u_0=5\times 10^8\rm cm~s^{-1}$ and $p_{max}$ between $10^3~m_p c$ and
$10^7~m_p c$ (different curves are labeled by the value of $\log_{10}(p_{max}/m_pc)$). 

The most striking feature in the behaviour of $R_{tot}$ versus $\xi$
is the rather sharp transition from strongly modified shocks (low
values of $\xi$, $R_{tot}\gg 4$) to weakly modified shocks (large
values of $\xi$, $R_{tot}\sim 4$). The even more striking point is
that, for $p_{max}>10^3 m_p c$,  in the very thin transition region 
in parameter space three solutions may appear. Multiple solutions
were initially found in two-fluid models \cite{dr_v80,dr_v81} and later
found in the models of \cite{malkov1,malkov2}. After introducing the
thermal leakage model for injection the appearance of multiple
solutions was found to be strikingly reduced and limited to very narrow
regions in parameter space (\cite{vannoni}). The multiple solutions
are here shown to exist even in Model B, though, again, in very narrow
regions of parameter space. In fact this region is so small that the
phenomenon was previously missed, since for standard values of the
parameters (in particular for $\xi=3.5$ which is most often used) we never
find more than one solution. 

Although we cannot prove it in a formal way at the present stage, it
is likely that the appearance of the multiple solutions in very narrow 
regions of parameter space is accompanied by the existence of
some type of instability that allows the system to chose among the three: 
from Fig.~\ref{fig:rtotxi} one can see that
the three solutions are found only in the region of values of $\xi$
for which the behaviour of the system suffers a transition from a
strongly modified shock ($\xi<\xi^*$) to a weakly modified shock 
($\xi>\xi^*$). In the transition region, a tiny change in the value of
$\xi$ around $\xi^*$ leads to either one or the other regime. It is
therefore likely that at least one of the solutions in this transition
region may be unstable. Of the three solutions that we find, one is
certainly non physical (the intermediate solution in
Fig. \ref{fig:rtotxi}) since it predicts that $R_{tot}$ increases with
increasing $\xi$.

The question of the stability of the other two solutions is still
open. Investigations on the stability of these shocks were carried out
by \cite{mond,topti,kangacoustic}.

The behaviour of $R_{tot}$ as a function of the Mach number is more
subtle in that one can change the Mach number by fixing the
temperature and varying the shock velocity or by fixing the shock
velocity and changing the temperature of the background gas, although
in terms of astrophysical applications the first case is probably the
most relevant or at least the most frequent. Below we consider the two
situations separately.  

\begin{figure}
\resizebox{\hsize}{!}{
\includegraphics{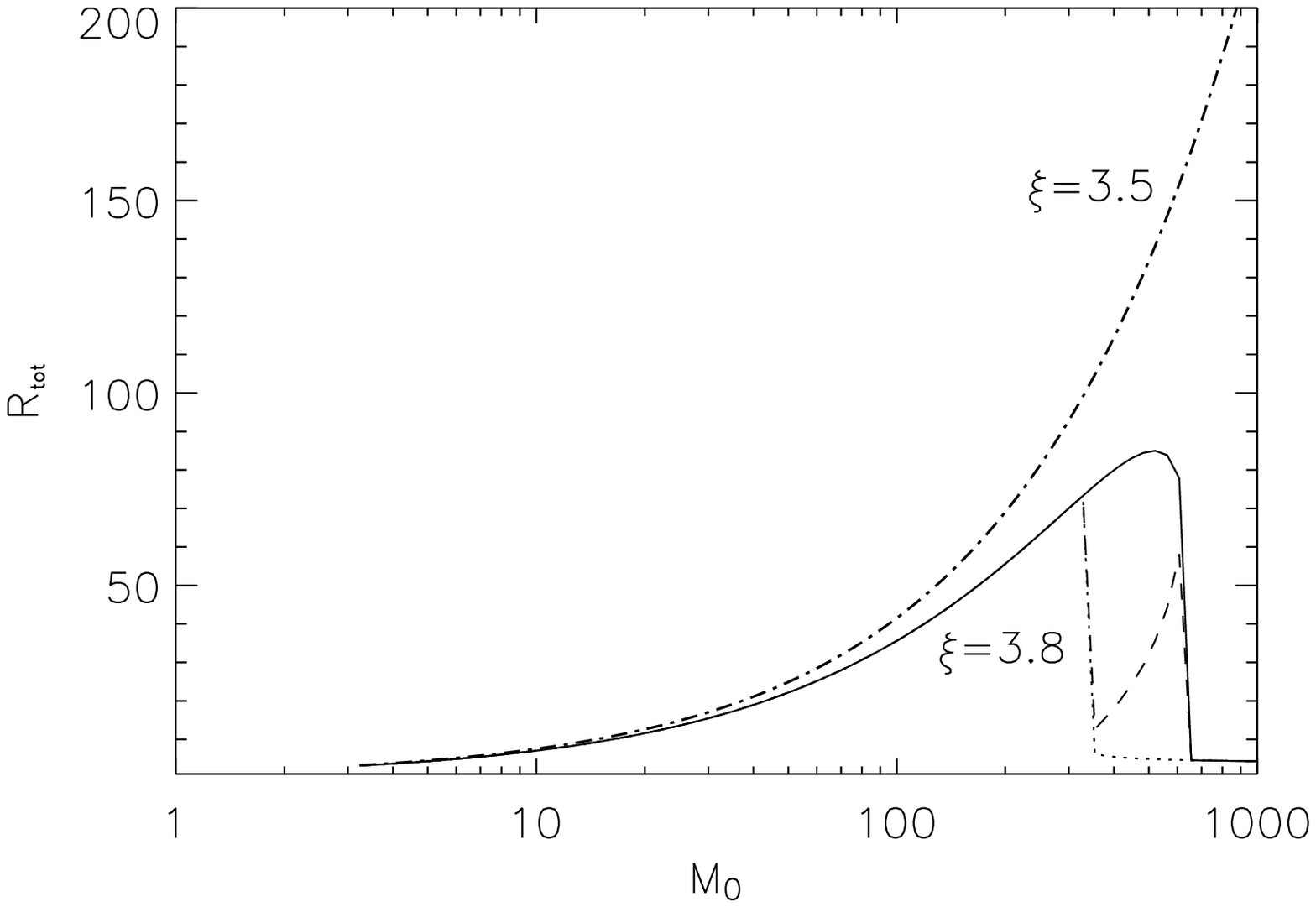}
\includegraphics{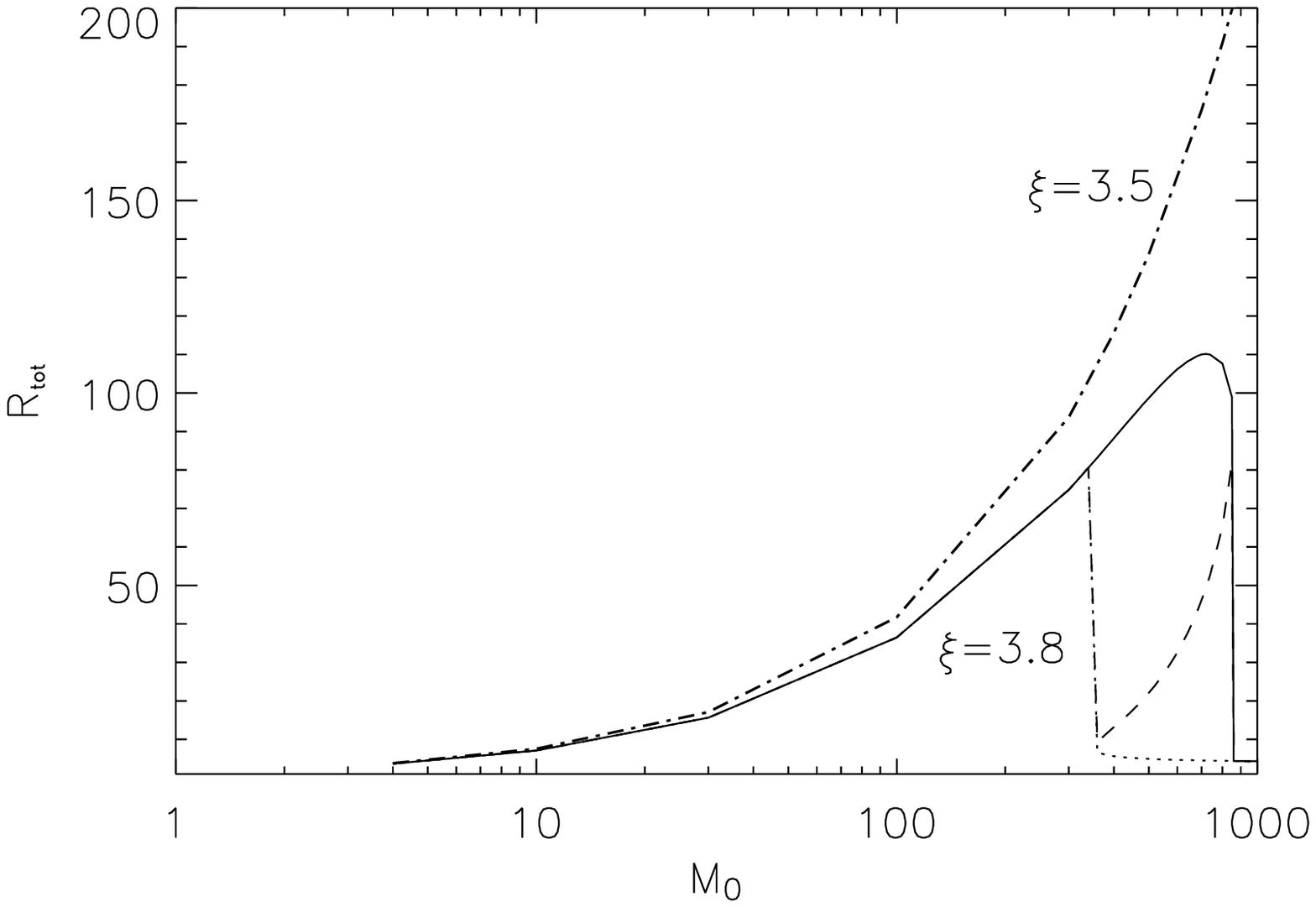}}
\caption{Total compression factor for the case at constant temperature
($T_0=10^4$ K) as a function of the Mach number. The left and right
  panel show the results of \citet{blasi1} and \citet{amato1}
  respectively. The two cases $\xi=3.5$ and $\xi=3.8$ are shown. 
  For $\xi=3.8$ multiple solutions appear. In this case the solid 
curve refers to the most modified one, the dotted curve to the one 
closer to linear and the dashed curve to the intermediate one.}
\label{fig:Tcost}
\end{figure}  

Increasing the Mach number (for a fixed temperature of the background
gas) leads to an increasing modification of the shock and therefore to
an increase in the value of $R_{tot}$. For $T_0=10^4$ K, $p_{max}=10^5
m_p c$ and $\xi=3.8$, the parameters used in the lower curves of 
Fig. \ref{fig:Tcost}, this trend continues up to $M_0\sim 500$.
For larger values of the Mach number, there is no
energy left to convert into accelerated particles and the shock
returns to be a {\it test particle} accelerator. In the thin
transition region between the strongly modified regime and the regime of 
weakly modified shocks
again three solutions appear. If $\xi=3.5$ is used instead 
of $\xi=3.8$ the transition moves to much larger Mach numbers,
of no astrophysical interest. These characteristics are shown
in Fig. \ref{fig:Tcost} for the model A (left panel) and 
B (right panel). The two sets of curves are in very good
agreement, although for the same parameters model B shows
the appearance of the multiple solutions at slightly larger Mach
numbers than the simple model A. The maximum value of the compression
factor is also slightly larger in the model of \cite{amato1} than it
is in the simple model. The upper curve in both plots refers to
$\xi=3.5$ and shows that no multiple solutions are found in this case.  

\begin{figure}
\resizebox{\hsize}{!}{
\includegraphics{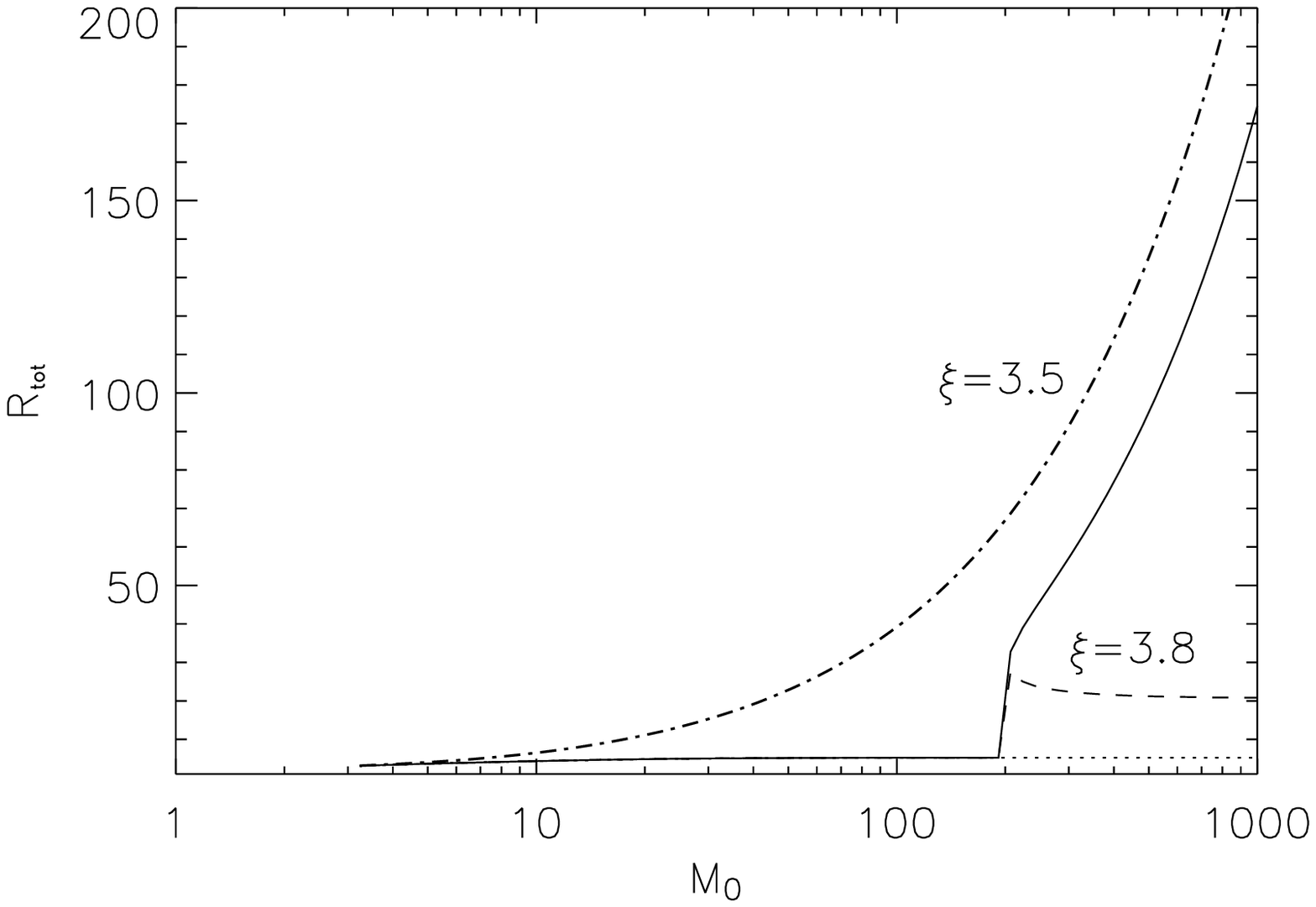}
\includegraphics{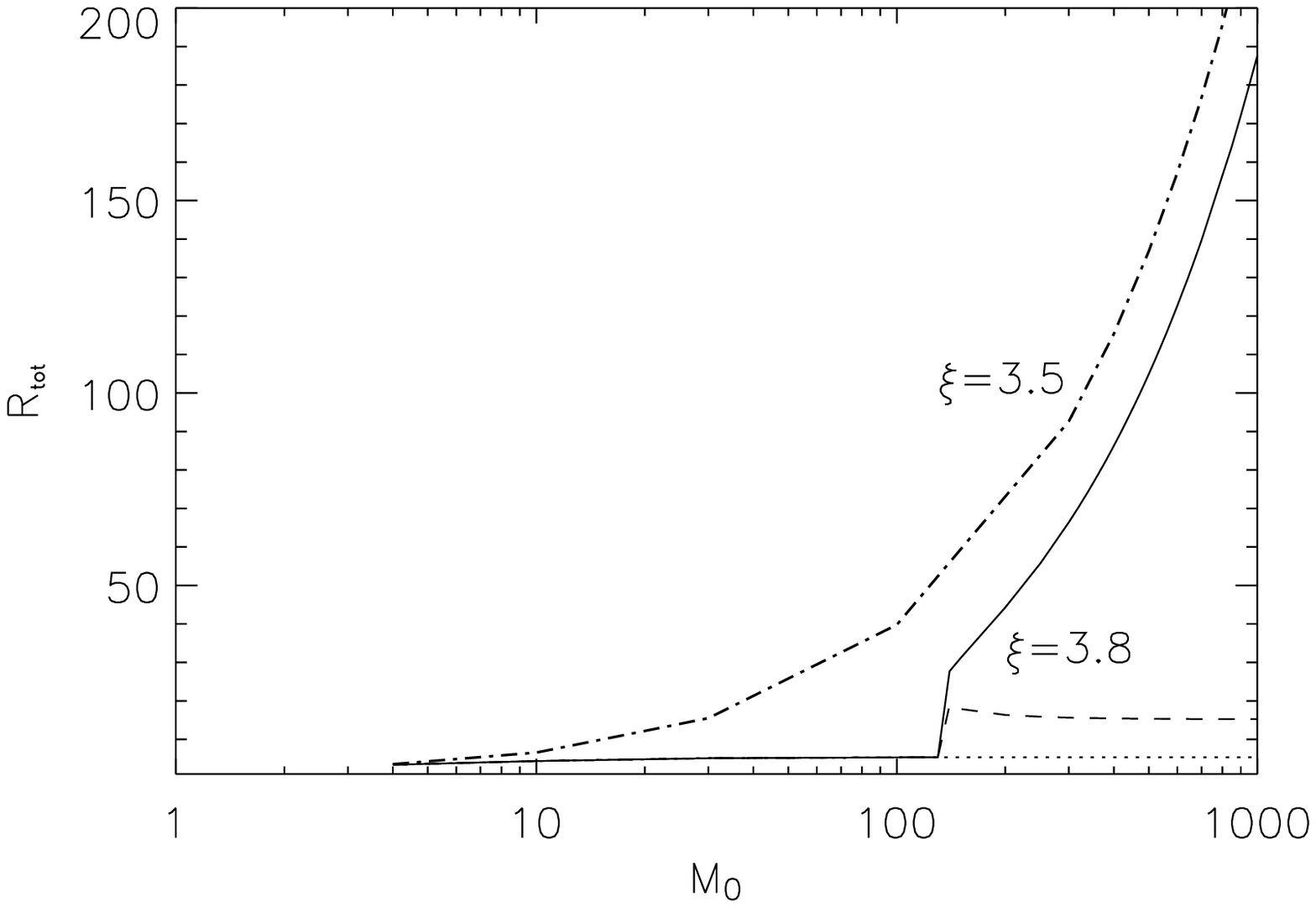}}
\caption{Total compression factor for the case at constant velocity 
  ($u_0=5\times 10^8\rm cm~s^{-1}$) as a function of the Mach number. The left 
  and right panel show the results of \citet{blasi1} and \citet{amato1}
  respectively. The two cases $\xi=3.5$ and $\xi=3.8$ are shown. 
  For $\xi=3.8$ multiple solutions appear. In this case the solid curve 
  refers to the most modified one, the dotted curve to the one closer to 
  linear and the dashed curve to the intermediate one.}
\label{fig:Ucost}
\end{figure}

When the increase of the Mach number is achieved by fixing the shock
velocity and changing the temperature, the behaviour of the total
compression factor as a function of $M_0$ is as shown in
Fig. \ref{fig:Ucost}. For $\xi=3.8$, oddly enough, the multiple
solutions appear for large values of $M_0$ and remain three
irrespective of how large $M_0$ becomes. At $M_0\sim 100-200$ (for the
values of the parameters adopted here) there is a bifurcation: one
branch that smoothly connects to the weakly modified solution for low
values of $M_0$ remains and tends asymptotically to $R_{tot}\sim 5$,
while two other branches appear, one with compression factor that
keeps increasing and the other with compression factor that tends to 
$R_{tot}\sim 15$. 

Even a qualitative comparison of Figs. \ref{fig:Tcost} and
\ref{fig:Ucost} reveals that the two cases are intrinsically different.
For $\xi=3.5$ (upper curve) no multiple solutions are found, and
  the shock always shows an appreciable level of modification.

The presence or absence of a transition to a weakly modified shock at
sufficiently large Mach number can be understood in a
semi-quantitative way by using the following argument. Let us assume
that the shock is weakly modified, namely that the pressure in the
form of accelerated particles is much smaller than $\rho_0 u_0^2$, so
that the spectrum can be approximated as a power law with slope $\sim
4$ ($f_0(p)\sim p^{-4}$). In this case the total pressure of the
accelerated particles at the shock location is 
\begin{equation}
P_{CR} = \frac{4\pi}{3}\int_{p_{inj}}^{p_{max}} dp p^3 v(p) f_0(p)
\approx
\frac{1}{3}c \left[ 4\pi p_{inj}^3 f_0(p_{inj}) \right]
p_{inj} \ln\left[\frac{p_{max}}{m_p c}\right].
\end{equation}
The term $4\pi p_{inj}^3 f_0(p_{inj})$ is roughly the total number of
particles at the shock location, which can be written as $\sim \eta n_1
u_1/u_2\approx 4\eta n_1$. It follows that
\begin{equation}
P_{CR} \approx
\frac{1}{3}c 4\eta n_1 p_{inj} \ln\left[\frac{p_{max}}{m_p c}\right].
\label{eq:limit}
\end{equation}
If one neglects the logarithmic term, the pressure in accelerated
particles is easily seen to scale with the injection momentum only.
In the context of the thermal leakage approach and in the limit
of strong shock, we can write $p_{inj}=\xi \sqrt{2 m_p k T_2}$, where
$k T_2=(3/16) u_1^2$. It follows that $p_{inj}=\xi \sqrt{(3/8)} m_p
u_1$, so that $P_{CR}\propto u_1$. The
linear scaling of the cosmic ray pressure with $u_1$ should be
compared with the ram pressure scaling, which is $\rho_1 u_1^2$. This
comparison immediately suggests that for sufficiently large values of
$u_1$ there is always enough kinetic pressure to fuel cosmic ray
acceleration without appreciable modification of the shock. Notice
that when the Mach number is increased by keeping the fluid velocity
constant, this argument does not apply and indeed in that case the
transition is not seen, while the three solutions persist at 
sufficiently high Mach numbers. If the Mach number is increased by increasing
the shock velocity (keeping the temperature constant) then the
transition is observed when the fluid velocity is larger (in fact
about one order of magnitude larger) than a critial value $u_*$, which 
can be estimated from Eq. \ref{eq:limit}:
\begin{equation}
\frac{u_*}{c} = \frac{8}{\sqrt{6\pi}} \xi^4 e^{-\xi^2}
\ln\left[\frac{p_{max}}{m_p c}\right]. 
\end{equation}
It is clear that the larger the values of $\xi$ the smaller the 
critical velocity for which the shock may become weakly modified.
The same threshold effect was previously found by \cite{simple}.

Some more discussion is required about the role of stationarity in the
appearance of multiple solutions. The phenomenon of multiple solutions
has been reported in the context of both two fluid models
\cite{dr_v80,dr_v81} and kinetic approaches, but in both cases stationarity
was assumed. Multiple solutions do not seem to appear in
time-dependent numerical calculations. From the physical point of view
a stationary solution of the diffusion-convection equation when
particle escape or energy losses are absent cannot exist. Even in the
context of the test-particle approach quasi-stationarity can be
recovered only at low momenta, far from $p_{max}$, while the temporal
evolution reflects into an increase of $p_{max}$. In the non-linear
regime an increase of $p_{max}$ leads to a modification of the
precursor and therefore of the entire spectrum. Hence the assumption
of stationarity is harder to justify in the context of a non-linear
theory of particle acceleration at shocks. This internal inconsistency
of both two-fluid models and semi-analytical kinetic models may be one
of the reasons why multiple solutions do not appear in time-dependent
approaches.

\subsection{Advected and escaping fluxes} 

One of the most important predictions of the non-linear theory of
diffusive particle acceleration at shocks is that for strongly
modified shocks an appreciable fraction of energy is in the form of
particles at the maximum momentum. On the other hand, these particles
are also the only ones that are allowed to leave the system from
upstream infinity. In other words, the return probability to the shock
from upstream is unity for all particles but for those with $p\sim
p_{max}$. The energy carried away from the shock by the highest energy
particles makes the shock radiative and the increased compressibility of
the background gas enhances the modification of the shock structure.
The equation for the conservation of the flux of energy
provides us with precious information, namely the flux of energy in
the form of accelerated particles that is advected towards downstream
infinity and the one that escapes towards upstream infinity. 

The conservation equation between downstream and upstream infinity can
be written in the following form: 

\begin{equation} 
\frac{1}{2} \rho_2 u_2^3 +\frac{\gamma_g}{\gamma_g-1} P_{g,2} u_2 +
\frac{\gamma_c}{\gamma_c-1} P_{c,2} u_2 =
\frac{1}{2} \rho_0 u_0^3 +\frac{\gamma_g}{\gamma_g-1} P_{g,0} u_0 - F_E,
\label{eq:energy}
\end{equation}
where $F_E$ is the flux of particles escaping at the maximum momentum from
the upstream section of the fluid (Berezhko \& Ellison, 1999). This
term is peculiar of modified shocks, being completely negligible when
acceleration takes place in the test particle regime, as we confirm in
the calculations below. 

In Eq. \ref{eq:energy} we can divide all terms by $(1/2)\rho_0 u_0^3$ and
calculate the normalized escaping flux:

\begin{equation}
F_E' = 1 - \frac{1}{R_{tot}^2} +\frac{2}{M_0^2 (\gamma_g-1)}
-\frac{2}{R_{tot}}\frac{\gamma_g}{\gamma_g-1}\frac{P_{g,2}}{\rho_0 u_0^2}
-\frac{2}{R_{tot}}\frac{\gamma_c}{\gamma_c-1}\frac{P_{c,2}}{\rho_0 u_0^2}.
\label{eq:energy_norm}
\end{equation}
From momentum conservation at the subshock we also have:
\begin{equation}
\frac{P_{c,2}}{\rho_0 u_0^2} = \frac{R_{sub}}{R_{tot}} -
\frac{1}{R_{tot}} + \frac{1}{\gamma_g M_0^2} \left( 
\frac{R_{sub}}{R_{tot}}\right)^{-\gamma_g},
\end{equation}
so that the escaping flux only depends upon the {\it environment}
parameters (for instance the Mach number at upstream infinity) and
the compression parameter $R_{sub}$ which is part of the solution.
The adiabatic index appropriate for cosmic rays, $\gamma_c$, is
here calculated self-consistently as:
\begin{equation}
\gamma_c = 1 + \frac{P_c}{E_c} = 1 + 
\frac{\frac{1}{3}\int_{p_{inj}}^{p_{max}} d p 4 \pi p^3 v(p) f_0(p)}
{\int_{p_{inj}}^{p_{max}} d p 4 \pi p^2 f_0(p) \epsilon(p)},
\end{equation}
where $E_c$ is the energy density in the form of accelerated particles
and $\epsilon(p)$ is the kinetic energy of a particle with momentum $p$.
It can be easily seen that $\gamma_c\to 4/3$ when the energy budget 
is dominated by the particles with $p\sim p_{max}$ (namely for strongly
modified shocks) and $\gamma_c\to 5/3$ for weakly modified shocks.
In Eq. \ref{eq:energy_norm} the term $F_{adv}'=\frac{2}{R_{tot}}
\frac{\gamma_c}{\gamma_c-1}\frac{P_{c,2}}{\rho_0 u_0^2}$ is clearly the 
fraction of flux which is advected downstream with the fluid. 

\begin{figure}
\resizebox{\hsize}{!}{
\includegraphics{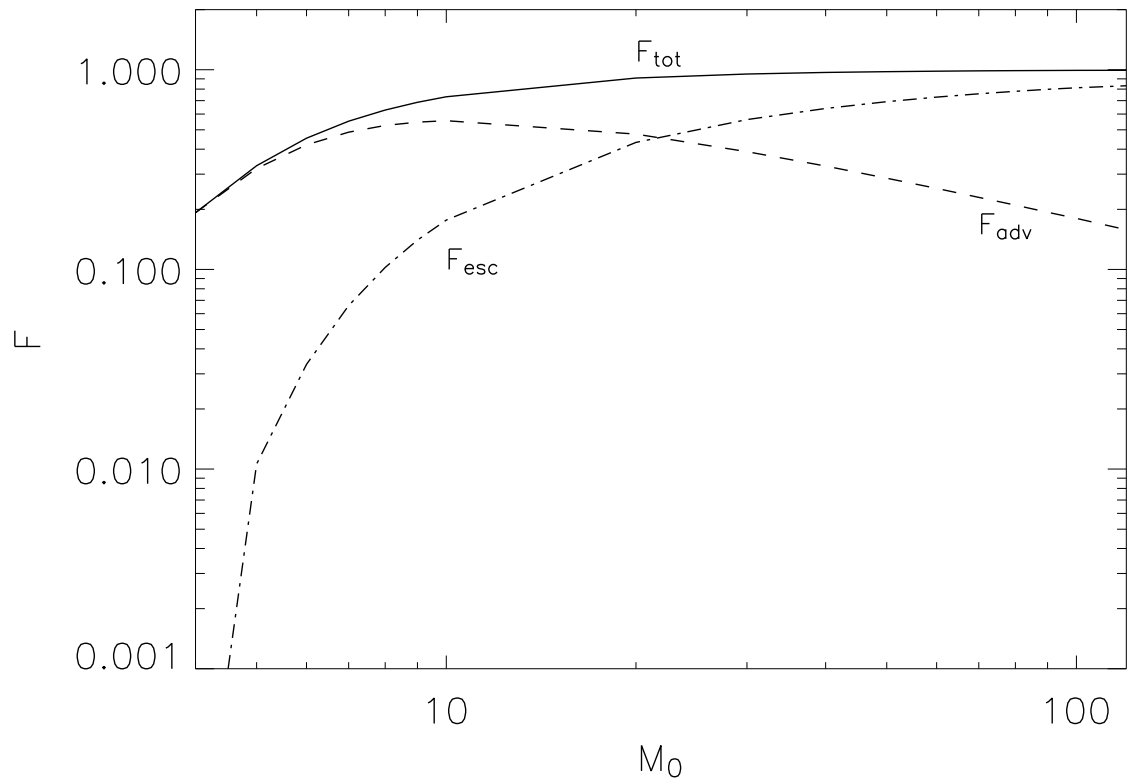}}
\caption{Normalized total (solid line), advected (dashed line) and
  escaping (dash-dotted line) flux of accelerated particles as
  functions of the Mach number.}
\label{fig:fluxes}
\end{figure}

The escaping flux, the advected flux and the total flux in the form of
accelerated particles are plotted in Fig. \ref{fig:fluxes} for 
$u_0=5\times 10^8 \rm cm~ s^{-1}$, $p_{max}=10^6~m_p c$ and
$\xi=3.5$. All fluxes are normalized to $(1/2)\rho_0 u_0^3$. Clearly
the difference between the total flux in accelerated particles and
unity gives the rate of conversion of the total energy into heating of
the background gas. For large Mach numbers this difference vanishes,
which implies that the gas is not appreciably heated at the shock, one
of the most impressive predictions of the non-linear theory of
particle acceleration.   

It is worth discussing in some details the physical meaning of the
advected and escaping fluxes. As we already pointed out earlier, the
escaping flux is all in the form of a narrow distribution in momentum
around $p\sim p_{max}$. One should keep in mind that in situations of
astrophysical interest, such as in the case of Supernova Remnants
(SNRs), after the beginning of the Sedov phase the maximum momentum
decreases with time. A distant observer is likely to observe (from a
single SNR) an overlap of peaked functions with $p\sim p_{max}(t)$. In
specific situations (e.g. \cite{ptuzira}) it has been shown that this
superposition leads to power law time-integrated spectra. Since the
particles of $p\sim p_{max}$ carry an appreciable fraction of energy
(in Fig. \ref{fig:fluxes} one can see that $F'_E\to 1$ for large Mach
numbers), this spectrum should roughly coincide with the {\it
injection} spectrum emitted by the SNR, despite the fact that the
spectrum at the shock has a shape of the type illustrated in
Fig. \ref{fig:spectrum} (at a given time). One should keep this
argument in mind when arguing about the effects induced by concavity
in the source spectra on the spectrum of diffuse cosmic rays observed
at the Earth. 

The spectrum of particles advected downstream, roughly speaking is
$\sim u_2 f_0(p)$, therefore the flux of energy carried by these
particles is  
$\sim u_2 P_c=\frac{1}{R_{tot}}P_c u_0$. Since for modified shocks
$R_{tot}\gg 1$, the actual advected flux is not very large. We
stress that this is the flux which is actually {\it useful} from the
phenomenological point of view in terms of conversion of energy into
gamma rays and other types of radiation within the source. On
the other hand the particles that escape from upstream may interact
with a thick target in the upstream region and produce a detectable
signal there (\cite{stefano,igor}). From Fig. \ref{fig:fluxes}
one can see that the advected flux takes less than $\sim 10\%$ of the
total energy influx for strongly modified shocks. Actually such flux
is larger for shocks which are less modified, which in
Fig. \ref{fig:fluxes} correspond to lower Mach numbers, but even in
this case it remains smaller than $\sim 30-40\%$ in units of
$(1/2)\rho_0 u_0^3$. 

The particles which are advected downstream remain behind the shock
and may eventually leave the remnant only at later times, suffering
adiabatic energy losses due to the expansion of the supernova shell. 
Their final spectrum, as could be observed by a distant observer, is a
complex convolution of the temporal evolution of the shell, the
changing maximum momentum, and the different levels of shock
modification at different times.

For low values of the Mach number the shock modification predicted by
kinetic models decreases, and as a consequence the concavity typical
of the modified spectra is reduced, until the test particle solution
is approached. Once the slope of the spectrum at $p\sim p_{max}$ drops
below $\sim 4$ the energy flux that escapes towards upstream infinity
gets suppressed, as shown in Fig. \ref{fig:fluxes}.

The physical origin of the escaping flux is again rather puzzling and
requires some comments: from the mathematical point of view the
requirement of having an escaping flux at upstream infinity derives
from imposing energy conservation. From the physical point of view it
is not obvious that the conditions for such escape exist. For
instance, in the case of supernova explosion, one may envision escape
of particles during the Sedov phase, but not during the free expansion
phase. In fact in the latter phase, the maximum momentum increases
with time. Nevertheless, a time-independent approach such as the
stationary kinetic models discussed here, would predict a flux escape
even during the free expansion phase, in order to conserve energy. 
Although this problem is usually not discussed in the
literature, we think that this may again call for the need to
develop time-dependent non-linear kinetic models. In such
approaches particle escape from upstream infinity should be invoked
only when there are the physical conditions for it to occur.

\subsection{The effect of turbulent heating on the energy fluxes}

In the previous section we have amply discussed that the non-linear
theory of particle acceleration tends to convert large fractions of
the flux traversing a shock into accelerated particles. One of the
effects that in astrophysical situations are likely to reduce such an
efficiency is the so-called turbulent heating. This generic expression
is used to refer to any process that may determine non-adiabatic gas
heating in the precursor. The two best known examples of this type of
processes are 
Alfv\`en heating (\cite{mv82}) and acoustic instability 
(\cite{drury}). Both effects are however very hard to implement
in a quantitative calculation: in the case of Alfv\`en heating, the
mechanism was originally introduced as a way to avoid the turbulent
magnetic field to grow to non-linear levels, while it is usually used
even in those cases in which $\delta B/B_0\gg 1$. Acoustic instability
develops in the pressure gradient induced by cosmic rays in the
precursor and results in the development of a train of shock waves
that heat the background gas (\cite{drury}). The analysis of
the instability is carried out in the linear regime, therefore it is
not easy to describe quantitatively the heating effect. In both cases
the net effect is the non-adiabatic heating of the gas in the
precursor, which results in the weakening of the precursor itself and
in the reduction of the acceleration effciency compared with the case
in which the turbulent heating is not taken into account. 

In order to simply illustrate the effect and compare the findings of 
model A and B in
describing the fluxes of advected and escaping accelerated particles
in the case of turbulent heating, we adopt the simple recipe provided
in \cite{simple} for Alfv\`en heating. The recipe consists in
modifying the relation 
between the gas pressure upstream of the subshock at the location $x$, 
$P_{g}(x)$, and the gas pressure at upstream infinity, in order to
take into account the amount of non adiabatic heating. The proposed
expression is  
\begin{equation}
\frac{P_{g}(x)}{P_{g,0}} = \left( 
\frac{\rho(x)}{\rho_0}
\right)^{\gamma_g} 
\left\{ 1 + (\gamma_g -1) \frac{M_0^2}{M_{A,0}}
  \left[ 1 - \left( \frac{\rho_0}{\rho(x)} \right)^{\gamma_g}
    \right]\right\}, 
\end{equation}
where $M_{A,0}$ is the Alfvenic Mach number at upstream infinity. One
can easily check that the non-adiabatic heating vanishes for
$M_{A,0}\to \infty$. With this simple recipe, Eq. \ref{eq:normalized1}
is modified in the following way:
\begin{equation}
\xi_c (x) = 1 + \frac{1}{\gamma_g M_0^2} - U(x) - \frac{1}{\gamma_g M_0^2}
U(x)^{-\gamma_g} \left\{ 1+(\gamma_g-1) \frac{M_0^2}{M_{A,0}} \left[
  1-U(x)^{\gamma_g} \right]\right\}.
\label{eq:pressmodified}
\end{equation}
If one keeps in mind that this modification also changes the relation
between $R_{sub}$ and $R_{tot}$, it is easy to realize that the 
computational procedures of the two kinetic models are left otherwise
unchanged and we are now able to determine the effect of turbulent
heating, at least in the context of this simple approach. 

\begin{figure}
\resizebox{\hsize}{!}{
\includegraphics{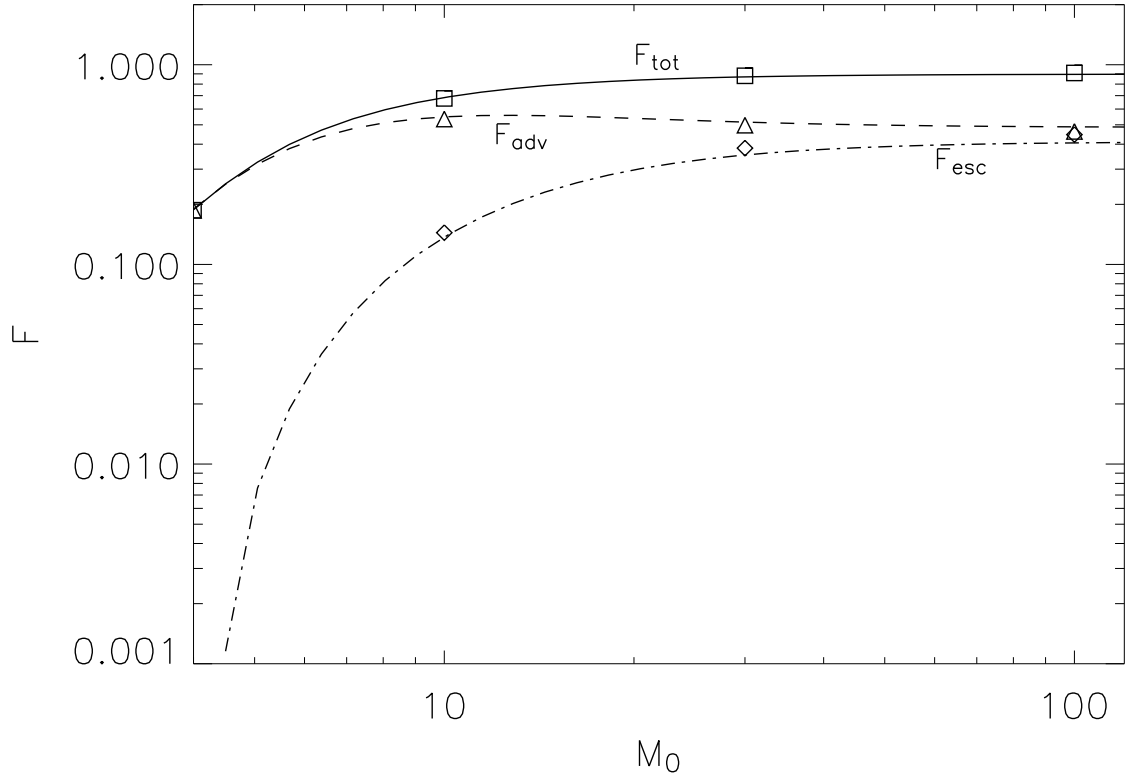}}
\caption{Normalized total (solid line), advected (dashed line) and escaping
  (dash-dotted line) flux of accelerated particles as functions of the
  Mach number when turbulent heating is taken into account. The assumed 
  value of the magnetic field is $B_0=10 \mu G$. The lines
  are obtained with the approach of \citet{blasi1}, while the symbols
  are the results of the method of \citet{amato1}.}
\label{fig:Hfluxes}
\end{figure}  

The best way to illustrate the effect of turbulent heating is by
plotting the same fluxes 
as in Fig. \ref{fig:fluxes} but including now the turbulent heating. 
The results are plotted in Fig. \ref{fig:Hfluxes}, where the lines are
obtained with Model A and the symbols represent the preditions of
Model B. Aside from the very good agreement
between the two methods, it is worth stressing that the turbulent
heating leads to less modified shocks, and therefore to a smaller flux
towards upstream infinity. The total flux (advected plus escaping) in
the form of accelerated particles is also substantially reduced, so
that appreciable heating can occur at the subshock. 

\section{The maximum momentum of the accelerated particles}\label{sec:pmax}

The maximum momentum that particles accelerated at modified shocks can
achieve has been calculated by \cite{caprioli}, for
arbitrary level of modification and for arbitrary choice of the
diffusion coefficient. Here we summarize the calculations of the
acceleration time and illustrate a recipe that allows us to include
this calculation in Model A. We compare the results to those obtained
with the formal approach of Model B. The acceleration time up to a
momentum $p$ is given by
\begin{equation}
<t> = - \left[\frac{\partial h}{\partial s}\right]_{s=0} = 
\frac{3 R_{tot}}{u_0^2}\
\int_{p_{inj}}^p \frac{dp'}{p'} \left\{
  \frac{R_{tot} D_2(p')}{R_{tot}U_p(p')-1} + \frac{u_0 \Lambda(p')}
{R_{tot} U_p(p') -1} \right\},
\label{eq:tacc}
\end{equation}
where $D_2(p)$ is the diffusion coefficient in the downstream plasma
and
\begin{equation} 
\Lambda(p)=\int_{-\infty}^0 dx \exp\left\{ \frac{q(p)}{3}
\left(1-\frac{u_2}{u_1}\right)
\int_0^x dx' \frac{u(x')}{D(x')}\right\}.
\label{eq:Lambda}
\end{equation}
In the limit of test particle acceleration, this acceleration time
reduces to the well known expression (\cite{lc83a,lc83b,drury83}):
\begin{equation}
<t> = \frac{3}{u_1 - u_2} \left[ \frac{D_2}{u_2} + \frac{D_1}{u_1}
\right].
\label{eq:lagage}
\end{equation}
We recall that in the model of \cite{blasi1} the information on the
spatial distributions in the precuror is kept only through the
relation $x(p)=D(p)/(u_0 U_p)$. This recipe has also been used to
couple $x$ to $p$ in the integral for $\Lambda(p)$. 

The maximum momentum is assumed here to be determined by the equality
between the acceleration time and the age of the accelerator. Clearly 
other recipes can be easily implemented as well. Below we consider two
cases: 1) a background magnetic field $B_0=10\mu G$; 2) the field is
amplified by streaming instability to the saturation value 
$\delta B=B_0 \left(2 M_{A,0} \frac{P_{CR}}{\rho_o u_0^2}\right)^{1/2}$
(\cite{bell78,amato2}). 

In both cases the diffusion coefficient is assumed to be Bohm-like. 
The magnetic field downstream, needed to calculate $D_2(p)$, remains 
$B_0$ in case 1), while in case 2) the two components of the field 
perpendicular to the shock normal are compressed by $R_{sub}$ whereas 
the parallel component is left unaltered. 

\begin{figure}
\resizebox{\hsize}{!}{
\includegraphics{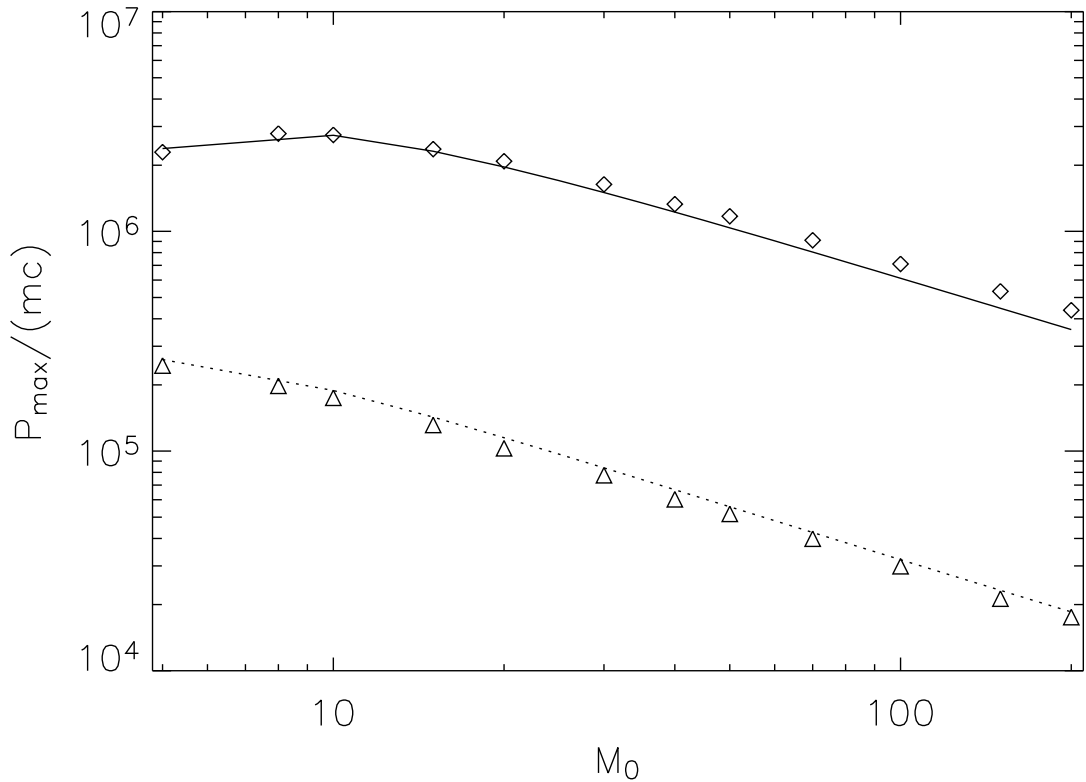}}
\caption{Maximum momentum of the accelerated particles for spatially
  constant diffusion coefficient. The lines show the result of the
  method of \citet{blasi1}, the symbols refer to (\citet{amato1}). The
  lower line (and symbols) refer to $B_0=10\mu G$, while the upper
  curve (and symbols) are obtained for magnetic field amplified by
  streaming instability. }
\label{fig:pmax}
\end{figure}

Our results for the maximum momentum as a function of the Mach number
are plotted in Fig. \ref{fig:pmax}, where we assumed that the
diffusion coefficient is spatially constant in the precursor. We also
adopted $u_0=5\times 10^8 \rm cm~s^{-1}$, $\xi=3.5$ and $B_0=10\mu
G$, while the age of the accelerator is fixed at $1000$ years. The 
curves in Fig. \ref{fig:pmax} refer to Model A, while the symbols
illustrate the results of \cite{caprioli} for Model B. The lower
curve (and symbols) refers to the case $B_0=10\mu G$, while the upper
curve (and symbols) refers to the case of magnetic field amplified by
streaming instability. Once again, the agreement between the two
kinetic approaches is very good, once Model A is completed with a
recipe that allows us to recover the spatial information in the
precursor.

As already noticed by \cite{caprioli}, the main factor in enhancing
the maximum momentum of the accelerated particles is the magnetic
field amplification. On the other hand, increasing the $p_{max}$ also
induces a larger shock modification (the same effect occurs when the
Mach number increases). Larger shock modifications lead to a slight
reduction of the maximum momentum. The reason is simple to understand:
as an order of magnitude estimate, the acceleration time is inversely
proportional to the square of a mean fluid velocity, as averaged over
the precursor. For strongly modified shocks the fluid upstream slows
down appreciably, thereby increasing the acceleration time. For our
benchmark values of the parameters, typical of supernova remnants, we
obtain maximum momenta between $5\times 10^5$ and $2\times 10^6~m_p
c$, comparable with the energy where the knee is observed in the cosmic
ray spectrum. 

\section{Conclusions}\label{sec:concl}

The recent growth of observational evidence of efficient particle
acceleration, at least in the case of supernova remnants, has posed a
serious challenge to describe the chain of physical processes that are all
taking place at the same time and in a strongly correlated manner:
particle acceleration is likely to excite streaming instability
leading to magnetic field amplification upstream. This field is then
advected downstream and leads to effective synchrotron emission of
electrons in narrow filaments which are observed in X-rays. 
The amplified magnetic field makes it possible to reach
maximum momenta of accelerated protons which are comparable with the
knee in the cosmic ray spectrum, but it also leads to, and requires,
efficient particle acceleration, which in turn modifies the shock
structure. The description of this complex chain of non-linear
phenomena ideally requires the particle acceleration process to be
treated at the same time of and coupled to a hydrodynamic (or even
MHD) code for the evolution of the remnant, thereby leading to the
need for a fast, efficient and accurate code for the description of
non-linear particle acceleration. In many previous papers
the simple approach of \cite{simple} was adopted. That
approach, though qualitatively appropriate, forced the spectrum of
accelerated particles to be a broken power law with given points where
the slopes changed, an assumption that can only be considered as a
{\it working hypothesis}. 

The approach of \cite{blasi1,blasi2}, discussed here as Model A,
is based on a physical ansatz on the spatial distribution of particles
in the precursor. It allows one to obtain the spectrum of accelerated
particles and all thermodynamical quantities of the fluid in a short 
computational time. For this reason it has recently been implemented
in the hydrodynamical code of \cite{noiSNR}. Model B, also discussed
here, leads to a formal solution of the problem of particle acceleration,
that we have shown to work well for a number of different 
assumptions on the spatial and momentum dependence of the diffusion 
coefficient. However its computational time is too long to allow one 
to use it in more complex calculations. 

In this paper we discussed the main results obtained by using the two 
kinetic approaches, with two goals in mind: 1) show that Model A
provides sufficiently accurate results to allow its use in more
complex calculations; 2) investigate one aspect of kinetic approaches
(and in fact of all stationary approaches)
that is still poorly understood, namely the appearance of multiple
solutions.     

In order to assess the goodness of Model A we compared its results
with the fomal solution of \cite{amato1,amato2} (Model B). This was
possible after completing Model A with a recipe to infer the spatial
information on particle distribution in the precursor, as discussed in
\S \ref{sec:spectra}. The differences between the results of the two
models are typically smaller than 20\% for all quantities which have
been calculated (spectra, velocity profile in the precursor,
compression factors). The recipe mentioned above also allowed us to
apply to Model A the calculation of the acceleration time as presented
by \cite{caprioli} and therefore to infer the maximum energy of the
accelerated particles. Also in this respect, as discussed in \S
\ref{sec:pmax}, Model A returns results in very good agreement with the
formal solution (Model B). 

The most ineteresting insights came however from the investigations on 
the presence of multiple solutions and escaping fluxes. 

Multiple solutions were first found in the context of stationary two
fluid models \cite{dr_v80,dr_v81} and later in the kinetic approach of
\cite{malkov1,malkov2}. \cite{vannoni} showed that treating injection
as a thermal leakage, multiple solutions persisted only in a very
narrow range of parameters. While confirming this finding here for
Model A, we found that also the formal solution of Model B leads to
multiple solutions, in about the same regions of parameters as for Model
A. Since the two methods are based on quite different criteria for the
solution of the equations, this is a strong hint to the fact that
the multiple solutions are an intrinsic property of the system and
not an artifact of the iteration procedure used to solve the
equations. A detailed discussion of the appearance of multiple
solutions has been presented in \S \ref{sec:multiple}. However a
physical understanding of the multiple solutions is still missing. It
is intriguing that they are found in time-independent approaches,
where stationarity is assumed, while they are not found in
time-dependent approaches. One should keep in mind that a stationary
solution of the equations for particle acceleration at a shock, in the 
absence of losses and escape does not exist, and that the assumption  
of stationarity is therefore rather artificial. 

On the other hand, time-dependent approaches are usually based on
the solution of the coupled transport equation and fluid equations in
such a way that the solution at a given time $t$ is advanced,
following a predefined integration scheme, to a time $t+\Delta t$. In
such approaches we cannot envision a procedure that would lead to the
appearance of multiple solutions. What probably could happen is that
there may be a strong dependence on initial conditions. To our
knowledge there has been no investigation of such effects. Moreover, 
as pointed out above, even if multiple solutions do exist one or more
of them are likely to be unstable. 

The requirement of stationarity implies that a flux of energy 
must escape from upstream infinity. Such a flux is actually predicted
on physical grounds in some circumstances, such as the slowing down of
the fluid motion in the Sedov phase of a SNR evolution, or because of
the fact that during such phase the magnetic field amplification is
expected to decrease with time, thereby reducing $p_{max}$. In this
way particles accelerated to higher maximum momentum at previous times
can no longer be confined in the accelerator and escape. But the
stationary non-linear calculations of particle acceleration at
modified shocks also predict escape in situations where it is not
immediate to foresee it on physical grounds. Again, since the role of
this escaping flux has profound implications on the acceleration
process it would be appropriate to investigate it using time-dependent 
techniques.

\end{document}